\documentclass[pre,aps,showpacs,byrevtex,amsmath,amssymb,twocolumn]{revtex4}
\usepackage{graphicx}

\begin{document}

\title{Statistical mechanics of triangulated ribbons}

\author{Boris Mergell}
\email{mergell@mpip-mainz.mpg.de}
\author{Mohammad R. Ejtehadi}
\author{Ralf Everaers}

\affiliation{Max-Planck-Institut f\"{u}r Polymerforschung,
 Postfach 3148, D-55021 Mainz, 
 Germany}

\date{\today}     

\begin{abstract}  
  We use computer simulations and scaling arguments to investigate
  statistical and structural properties of a semiflexible ribbon
  composed of isosceles triangles. We study two different models, one
  where the bending energy is calculated from the angles between the
  normal vectors of adjacent triangles, the second where the edges are
  viewed as semiflexible polymers so that the bending energy is
  related to the angles between the tangent vectors of next-nearest
  neighbor triangles. The first model can be solved exactly whereas
  the second is more involved. It was recently introduced by Liverpool
  and Golestanian Phys.Rev.Lett. {\bf 80}, 405 (1998),
  Phys.Rev.E {\bf 62}, 5488 (2000) as a model for
  double-stranded biopolymers such as DNA.  Comparing observables such
  as the autocorrelation functions of the tangent vectors and the
  bond-director field, the probability distribution functions of the
  end-to-end distance, and the mean squared twist we confirm the
  existence of local twist correlation, but find no indications for
  other predicted features such as twist-stretch coupling, kinks, or
  oscillations in the autocorrelation function of the bond-director
  field.
\end{abstract}


\maketitle   


\section{Introduction}
\label{section:intro}

A characteristic feature of many biopolymers is their high bending
stiffness. Contour lengths of the order of $\mu m$ and persistence
lengths of the order of $50nm$ in the case of DNA even allow
microscopy techniques to be used to directly observe their structure
and dynamics~\cite{Perkins_sci_95,Perkins_sci_97}. The model mostly
used to interpret recent experimental data of micromechanical
manipulations of single DNA
chains~\cite{Bustamente_sci_95,Perkins_sci_95,Perkins_sci_97,Strick_sci_96,Strick_pnas_98,Cluzel_sci_96}
is that of the Kratky-Porod wormlike chain in which the polymer
flexibility is determined by a single length, the persistence length
$l_p$. Generalizations account for the chain helicity and coupling
terms between bending, stretching, and twisting allowed by
symmetry~\cite{Yamakawa,MarkoSiggia_mm_95,MarkoSiggia_mm_94,MarkoSiggia_pre_95,MarkoSiggia_sci_95,Moroz_pnas_97,Kamien_epl_97,Rabin_condmat_01,Rabin_pre_00,Rabin_prl_00,KehrbaumMaddocks_ptrsl_97,ManningMaddocks_jcp_96}.
All these continuum models of DNA neglect the double-stranded
structure of DNA and one may ask, if this feature could not cause
qualitative different behavior.

The bending stiffness of single- and double-stranded DNA, for example,
differs by a factor of 25~\cite{Kamenetskii_90}.  The simplest model
which takes the double-strandedness into account is the railway-track
model~\cite{Everaers_epl_95} where two wormlike chains are coupled
with harmonic springs. In two dimensions one finds drastical
consequences: the bending fluctuations in the plane of the ribbon are
strongly suppressed. The molecule becomes effectively stiffer on
larger length scales. But the relevant question is: what are the
effects in three dimensions? Liverpool {\em et
  al}~\cite{Liverpool_prl_98,LiverpoolGolestian_pre_00} investigated a
version of the railway-track model in three dimensions where bending
in the plane of the ribbon is forbidden by a constraint. Using
analytical and simulation techniques they predict the existence of a
low temperature regime where ribbons adopt a kink-rod structure due to
a spontaneously appearing short-range twist structure resulting in an
oscillatory behavior of the autocorrelation function of the
bond-director field.  Furthermore a twist-stretch coupling is
predicted.

We study the discretized version of the simulation model of Liverpool
{\em et al}~\cite{Liverpool_prl_98,LiverpoolGolestian_pre_00} in the
low temperature regime with the help of scaling arguments and MC
simulations. In order to understand and to quantify the effects
arising from the local twist structure of the Liverpool model we
compare it with an analytically more tractable model where the bending
stiffness is defined via the interaction of the normal vectors so that
there is no tendency to helical structures.  Furthermore, we perform
several MC simulation runs with an additional external force in order
to test if the preferred buckling mechanism occurs via kinks.

\section{Continuous description of two coupled semiflexible chains}

A ribbon is an inextensible, unshearable rod which can be
parameterized by the arclength $s$. To each point $s$ one attaches a
triad of unit vectors $\{{\mathbf d}_i(s)\}$. The vectors ${\mathbf
  d}_1(s)$ and ${\mathbf d}_2(s)$ are directed along the two principle
axis of the cross section, the vector ${\mathbf d}_3(s)$ is the
tangent vector. As the triad is an orthonormal basis set they satisfy
kinematic equations of the form
\begin{equation}
\label{frenet_general}
\frac{d}{ds}{\mathbf d}_i(s)=\epsilon_{ijk}u_j(s){\mathbf d}_k(s)
\end{equation}
with $\epsilon_{ijk}$ being the alternating tensor and $u_j(s)$
representing bend ($u_1(s)$ out-of-plane, and $u_2(s)$ in-plane) and
twist strains ($u_3(s)$) respectively. One can find a relation between
the ordinary Frenet equations containing only two parameters, the
curvature $\kappa(s)$ and the torsion $\tau(s)$
\begin{eqnarray}
\frac{d{\mathbf t}(s)}{ds} &=& \kappa(s){\mathbf n}(s)\\
\frac{d{\mathbf b}(s)}{ds} &=& -\tau(s){\mathbf n}(s)\\
\frac{d{\mathbf n}(s)}{ds} &=& \tau(s){\mathbf b}(s)-\kappa(s){\mathbf t}(s) 
\end{eqnarray}
and Eqs.~(\ref{frenet_general}) by fixing ${\mathbf d}_3(s)={\mathbf
  t}(s)$ so that ${\mathbf d}_1(s)$ and ${\mathbf d}_2(s)$ are given
by a rotation around ${\mathbf t}(s)$ with angle $\Psi(s)$. In this
context $\Psi(s)$ can be seen as the twist
angle~\cite{Rabin_pre_00,Maggs_jcp_01}. A straightforward calculation
gives for the generalized torsions: $u_1(s)=-\frac{d}{ds}{\mathbf
  d}_3(s)\cdot{\mathbf d}_2(s)=\kappa(s)\cos\Psi(s)$,
$u_2(s)=\frac{d}{ds}{\mathbf d}_3(s)\cdot{\mathbf
  d}_1(s)=\kappa(s)\sin\Psi(s)$, and $u_3(s)=\frac{d}{ds}{\mathbf
  d}_1(s)\cdot{\mathbf d}_2(s)=\tau(s)+\frac{d\Psi(s)}{ds}$. The total
Twist $Tw$ of a ribbon is thus given by the integration of the local
twist $u_3(s)$ along the contour normalized by the factor $2\pi$
\begin{equation}
\label{def:twist}
Tw=\frac{1}{2\pi}\int_0^L\,u_3(s)ds
\end{equation}
with $L$ being the contour length.
Together with the parameter set $\hat{u}_i(s)$, which determines
whether the stress-free reference configuration includes spontaneous
curvature and twist, the elastic part of the Hamiltonian is usually
defined by quadratic terms in
$u_i(s)-\hat{u}_i(s)$~\cite{Rabin_condmat_01,Rabin_pre_00,Rabin_prl_00,MarkoSiggia_mm_94,MarkoSiggia_pre_95,MarkoSiggia_sci_95,Moroz_pnas_97,Kamien_epl_97,KehrbaumMaddocks_ptrsl_97,ManningMaddocks_jcp_96,Joanny_jp2_96}.

It is an interesting question to which extent this generic description 
applies to more microscopic models of DNA~\cite{Lai_cpl_01}. The simplest case is that
of a ``railway track'' or ladder model consisting of two (or more)
semiflexible chains 
\begin{equation}  
\label{h_semi}
{\cal H}_{tt}=\frac{k}{2}\int_0^L\,ds\left\{\left(\frac{d^2{\mathbf
        r}_1(s)}{ds^2}\right)^2 + \left(\frac{d^2{\mathbf
        r_2}(s)}{ds^2}\right)^2\right\},
\end{equation}
plus a coupling between opposite points on different
chains~\cite{Everaers_epl_95}. Liverpool {\em et
  al}~\cite{Liverpool_prl_98,LiverpoolGolestian_pre_00} considered the
limit where the distance $a$ between the coupling points (i.e. the
width of the ribbon) is imposed as a rigid constraint which prevents
bending in the plane of the ribbon:
$\frac{d{\mathbf{t}}(s)}{ds}\cdot{\mathbf{b}}(s)=0$ where
${\mathbf{t}}(s)=\frac{d{\mathbf{r}}(s)}{ds}$ is the tangent vector to
the mid-curve
${\mathbf{r}}(s)={\mathbf{r}}_1(s)-\frac{a}{2}={\mathbf{r}}_2(s)+\frac{a}{2}$
and ${\mathbf{b}}(s)$ is the bond-director pointing from one strand to
the other. Note, that the constraint is equivalent to $\Psi(s)=0$.
Rewriting Eq.~(\ref{h_semi}) in terms of ribbon variables they found
\begin{equation}  
  {\cal H}_{tt} = \frac{k}{2}\int_0^L\,ds\left\{2\left(\frac{d^2{\mathbf
          r}(s)}{ds^2}\right)^2 + \frac{a^2}{2}\left(\frac{d^2{\mathbf
          b}(s)}{ds^2}\right)^2\right\}
\end{equation}
which can also be expressed as
\begin{eqnarray}
  \left(\frac{d{\mathbf{t}}}{ds}\right)^2 &=& \kappa^2\\
  \left(\frac{d^2{\mathbf b}}{ds^2}\right)^2 &=& \left(\frac{du_1}{ds}\right)^2 +
  \left(u_1^2-u_3^2\right)^2 + \left(\frac{du_3}{ds}\right)^2\nonumber\\
  &=& \left(\frac{d\kappa}{ds}\right)^2 +
  \left(\frac{d\tau}{ds}\right)^2 + \left(\kappa^2-\tau^2\right)^2.
\end{eqnarray}
Note, that henceforth we use ${\mathbf{b}}(s)$ as the bond-director and
${\mathbf{n}}(s)$ as the normal vector to the ribbon plane.


\section{Geometry of triangulated ribbons}
\label{section:model}

\begin{figure}[t]  
  \begin{center}
    \includegraphics[angle=0,width=0.99\linewidth]{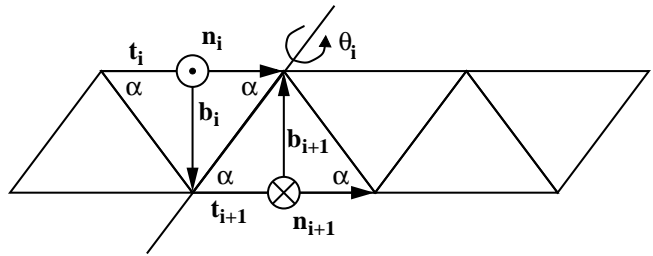}
  \end{center} 
  \caption{Illustration of the used variables. The length of each
    triangle $|{\mathbf{t_i}}|$ corresponds to the bond length $b$ and
    the height $|{\mathbf{b_i}}|=\frac{1}{2}b\tan(\alpha)$ defines the
    strand separation length. $\{\theta_i\}$ term the folding angles.}
  \label{fig_iso}
\end{figure}

Following Liverpool {\em et
  al}~\cite{Liverpool_prl_98,LiverpoolGolestian_pre_00} we consider
ribbons discretized by triangulation. In order to extract some
fundamental properties of double-stranded semiflexible polymers we
consider a ribbon-like system composed of isosceles triangles as shown
in Fig.~\ref{fig_iso}. The orientation of each triangle is given by
$N-1$ rotations around the edges of the triangles with folding angles
$\{\theta_i\}$.  $N$ is the number of triangles characterized by a set
of trihedrons $\{{\mathbf{t_i}},{\mathbf{b_i}},{\mathbf{n_i}}\}$ where
${\mathbf{t_i}}$ is the tangent vector of the $i$th triangle,
${\mathbf{b_i}}$ is the bond-director, and ${\mathbf{n_i}}$ is the
normal vector.
Going from one set of trihedrons
$\{{\mathbf{t_i}},{\mathbf{b_i}},{\mathbf{n_i}}\}$ to the neighbor set
$\{{\mathbf{t_{i+1}}},{\mathbf{b_{i+1}}},{\mathbf{n_{i+1}}}\}$ implies
a rotation ${\cal R}_i$ around the edge between the respective
triangles with angle $\theta_i$ and a reflection of ${\mathbf{b_i}}$
and ${\mathbf{n_i}}$, i.e.
\begin{equation}
  \label{eq:1}
  \begin{pmatrix} {\mathbf{t_{i+1}}} \\ {\mathbf{b_{i+1}}} \\ {\mathbf{n_{i+1}}} \end{pmatrix} 
  = {\cal T} {\cal R}_i \begin{pmatrix} \mathbf{t_{i}} \\ \mathbf{b_{i}} \\ \mathbf{n_{i}} \end{pmatrix}
\end{equation}
with
\begin{eqnarray}
  \label{eq:2}
  {\cal T} &=&
  \begin{pmatrix} 1 & 0 & 0 \\ 0 & -1 & 0 \\ 0 & 0 & -1 \end{pmatrix} \\
  \label{eq:3}
  {\cal R}_i &=&
  \begin{pmatrix} {\mathbf{t_i}}\cdot{\mathbf{t_{i+1}}} & {\mathbf{t_i}}\cdot{\mathbf{b_{i+1}}} &
    {\mathbf{t_i}}\cdot{\mathbf{n_{i+1}}} \\
    {\mathbf{b_i}}\cdot{\mathbf{t_{i+1}}} &
    {\mathbf{b_i}}\cdot{\mathbf{b_{i+1}}} &
    {\mathbf{b_i}}\cdot{\mathbf{n_{i+1}}} \\
    {\mathbf{n_i}}\cdot{\mathbf{t_{i+1}}} &
    {\mathbf{n_i}}\cdot{\mathbf{b_{i+1}}} &
    {\mathbf{n_i}}\cdot{\mathbf{n_{i+1}}} \end{pmatrix}.
\end{eqnarray}
The matrix product ${\cal T} {\cal R}_i$ can be viewed as a transfer
matrix. The evaluation of the scalar products of ${\cal R}_i$ gives
\begin{equation}
  \label{eq:4}
  \begin{split}
    {\cal R}_{i,11} &= \cos(\theta_i) +
    \cos(\alpha)^2\left(1-\cos(\theta_i)\right) \\
    {\cal R}_{i,12} &=
    -\cos(\alpha)\sin(\alpha)\left(1-\cos(\theta_i)\right) \\
    {\cal R}_{i,13} &= -\sin(\alpha)\sin(\theta_i) \\
    {\cal R}_{i,21} &=
    \cos(\alpha)\sin(\alpha)\left(1-\cos(\theta_i)\right) \\
    {\cal R}_{i,22} &= \cos(\theta_i) +
    \sin(\alpha)^2\left(1-\cos(\theta_i)\right) \\
    {\cal R}_{i,23} &= -\cos(\alpha)\sin(\theta_i) \\
    {\cal R}_{i,31} &= \sin(\alpha)\sin(\theta_i) \\
    {\cal R}_{i,32} &= \cos(\alpha)\sin(\theta_i) \\
    {\cal R}_{i,33} &= \cos(\theta_i).
  \end{split}
\end{equation}

\begin{figure}[t]  
  \begin{center}
    \includegraphics[angle=0,width=0.88\linewidth]{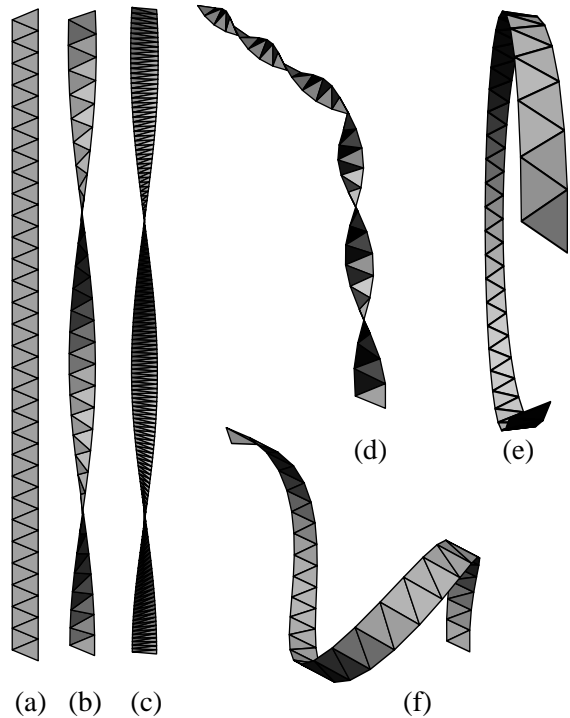}
  \end{center} 
  \caption{Illustration of bending, twisting, and kinking. (a) A flat
    ribbon as ground state conformation. (b) A twisted structure (c)
    The same twisted structure obtained with a smoother
    discretization. (d) Unlike twists meeting at the center resulting
    in a kink with $\theta_i$ positive for $i<N/2$, negative for $i\ge
    N/2$, and $|\theta_i|=|\theta_{i+1}|$, i.e.
    $\delta\theta_i=0,\,\forall i\ne N/2$ and
    $\delta\theta_{N/2}=2\theta_{N/2}$. (e) A bent structure. (f) A
    mixture of bent and twist resembling a solenoidal structure.}
  \label{fig:TwBe}
\end{figure}

In order to quantify properties such as bending and twisting within
the given discretization we study the relation between the folding
angles $\theta_i$ and these quantities which is illustrated in
Fig.~\ref{fig:TwBe}. One recognizes that the chain is not bent in case
of $\theta_i-\theta_{i+1}=\delta\theta_i=0$ and that one gains purely
twisted structures if $\theta_i\equiv const$. On the other hand the
chain is untwisted but bent if $\delta\theta_i=2\theta_i$. In case of
$\theta_i\ne\pm\theta_{i+1}$ and $\theta_i\ne0$ the chain is bent and
twisted simultaneously resulting in solenoidal/torsional structures as
is illustrated in Fig.~\ref{fig:TwBe}(f). A kink is characterized by
unlike twists meeting at an edge as it is shown in
Fig.~\ref{fig:TwBe}(d).

Due to the triangulation of the ribbon one has to consider three
triangles to calculate the discretized expressions for the
out-of-plane bending strain $u_1(s) =-\frac{d}{ds}{\mathbf
  t}(s)\cdot{\mathbf n}(s) \approx -\frac{{\mathbf t}(s+\Delta
  s)-{\mathbf t}(s)}{\Delta s}\cdot{\mathbf n}(s) = -\frac{1}{\Delta
  s}{\mathbf t}(s+\Delta s)\cdot{\mathbf n}(s)$ and the twist strain
$u_3(s) =\frac{d}{ds}{\mathbf b}(s)\cdot{\mathbf n}(s) \approx
\frac{{\mathbf b}(s+\Delta s)-{\mathbf b}(s)}{\Delta s}\cdot{\mathbf
  n}(s) = \frac{1}{\Delta s}{\mathbf b}(s+\Delta s)\cdot{\mathbf
  n}(s)$ which we call $\kappa_i$ and $\tau_i$ respectively. The local
curvature $\kappa_i$ and the local twist rate $\tau_i$ between
triangle $i$ and $i+2$ are therefore given by
\begin{eqnarray}
  \label{kappa}
  \kappa_{i} &\equiv&
  -\frac{1}{b}\sum_{j=i}^{i+1}{\mathbf{n}_j}\cdot{\mathbf{t}_{j+1}}\approx\frac{\sin(\alpha)}{b}\delta\theta_i\\
  \label{tau}
  \tau_{i} &\equiv& \frac{1}{b}\sum_{j=i}^{i+1}{\mathbf{n}_j}\cdot{\mathbf{b}_{j+1}}\approx
  \frac{\cos(\alpha)}{b}(\theta_i+\theta_{i+1}),
\end{eqnarray}
where the accuracy of the right-hand side expressions only depends on
the refinement of the discretization, i.e. on the values of $b$ and
$\alpha$.
Hence a spontaneous bending can be introduced via an additional term
to the Hamiltonian with ${\cal H}_{curv} = k_{curv}
\sum_i\left(\sum_{j=i}^{i+1}{\mathbf{n}_j}\cdot{\mathbf{t}_{j+1}} -
  \delta\theta_{sp,i}\right)^2$ and a spontaneous twist can be
introduced by an additional term ${\cal H}_{Tw} = k_{Tw}
\sum_i\left(\sum_{j=i}^{i+1}{\mathbf{n}_j}\cdot{\mathbf{b}_{j+1}} -
  \theta_{sp,i}\right)^2$.  Note, that the total twist $Tw$ is given by
$Tw=1/(2\pi)\sum_i\tau_i$.

\section{Model description}

The bending stiffness within the given discretization can be taken
into account by various interactions. One possible definition of a
bending stiffness, which makes the problem analytically tractable, is a
nearest neighbor interaction (plaquette stiffness) between the normal
vectors $\{{\mathbf{n_i}}\}$ in analogy to the triangulation of
vesicles~\cite{Gompper_pra_92} which results in the following
Hamiltonian
\begin{equation}
  \label{fullhn}
  \frac{{\cal H}_{nn}}{k_BT} = 
  k\sum_{i=1}^{N-1}\left(1+{\mathbf{n_i}}\cdot{\mathbf{n_{i+1}}}\right).
\end{equation} 
In contrast Liverpool {\em et
  al}~\cite{Liverpool_prl_98,LiverpoolGolestian_pre_00} were
interested in the statistical mechanics of coupled wormlike chains and
therefore chose a next-nearest neighbor interaction (edge stiffness)
between the tangent vectors $\{{\mathbf{t_i}}\}$ with rigidity $k$ so
that the Hamiltonian is given by
\begin{equation} 
  \label{fullht}
  \frac{{\cal H}_{tt}}{k_BT} =
  k\sum_{i=1}^{N-2}\left(1-{\mathbf{t_i}}\cdot{\mathbf{t_{i+2}}}\right).
\end{equation}
Both definitions lead to a flat ribbon as the ground state conformation for
zero temperatures $T=0$. 

The above defined interactions lead to very distinct conformational
features of the ribbon which can be understood by building up the
ribbon just by adding successively the triangles in the absence of
thermal fluctuations. Assuming that $\theta_1\ne0$ all subsequent
angles $\theta_i$ with $i>1$ vanish in the case of the nearest
neighbor interaction (${\cal H}_{nn}$). In contrast the
tangent-tangent interaction (${\cal H}_{tt}$) leads to the formation
of a helix with $\theta_i=\theta_{i+1}$ as a result of the enforced
alignment of the tangent vectors. This suggests a correlation of the
folding angles $\{\theta_i\}$ which entails at least locally helical
structures.

Assuming that the chains are rather stiff (continuum limit), i.e.
small folding angles $\theta_i$, one can expand the Hamiltonians with
regard to $\theta_i$. Since ${\cal H}_{nn}$ is diagonal in $\theta_i$,
it is sufficient to consider terms up to second order. ${\cal H}_{tt}$
contains coupling terms between $\theta_i$ and $\theta_{i+1}$ which
makes it necessary to keep terms up to fourth order in the analysis:
\begin{eqnarray}
  \label{hn}
  \frac{{\cal H}_{nn}}{k_BT} &\approx& \frac{k}{2}\sum_{i=1}^{N-1}\theta_i^2\\
  \label{ht}
  \frac{{\cal H}_{tt}}{k_BT} &\approx& \frac{k}{2}\sum_{i=1}^{N-2}
  \left\{\sin(\alpha)^2\delta\theta_i^2
  \left(1-\frac{1}{12}\delta\theta_i^2\right)\right. \notag\\
  &+& \left.
  \sin(\alpha)^2\cos(\alpha)^2\theta_i^2\theta_{i+1}^2\right\}
\end{eqnarray}
with $\delta\theta_i=\theta_i-\theta_{i+1}$.


\section{MC Simulation}

Both models have local interactions and can be studied conveniently
using a dynamic MC scheme. Trial moves consist of small random changes
of the folding angles by a small amplitude $1/\sqrt{k}$, where $k$ is
the bending stiffness, and are accepted or rejected according to the
Metropolis scheme~\cite{Metropolis_jcp_53}. In the simulations we
always use the full Hamiltonians Eq.~(\ref{fullhn})
and~(\ref{fullht}). MC moves changing the folding angles correspond to
the well-known Pivot algorithm~\cite{Binder_00}. The conformations are
subsequently recalculated from Eqs.~(\ref{eq:1})-(\ref{eq:4}) and
analyzed. Each simulation run comprises $100000$ MC-moves where one MC
move corresponds to $N-1$ trials with $N$ being the number of
triangles. The longest correlation time we observed was of the order
of 50 MC moves for the total twist of the chain. In order to check if
equilibrium is reached we compared simulation runs with a flat initial
conformation, i.e $\theta_i=0$, with simulation runs with crumpled
conformations corresponding to equally distributed angles $\theta_i$
out of the interval $[-1/\sqrt{k};1/\sqrt{k}]$. Both runs yield the
same results for the calculated observables.


\section{Plaquette Stiffness}
\label{section:A}

Since the Hamiltonian ${\cal H}_{nn}$ of Eq.~(\ref{hn}) is quadratic
and diagonal in $\theta_i$ the solution in angle space is trivial. As
a consequence of the independence of successive folding angles it
yields $\langle\theta_i\theta_j\rangle=\frac{1}{k}\delta_{ij}$ and
$\langle{\cal A}\rangle = \langle\prod_{k=i}^{j}({\cal T}{\cal
  R}_k)\rangle = \langle{\cal T}{\cal R}_k\rangle^{j-i}$ where the
matrix product is carried out in the eigenvector basis of
$\langle{\cal T}{\cal R}_k\rangle$ (the eigenvectors depend only on
the geometry of the triangles). The diagonal elements of $\langle{\cal
  A}\rangle$ are the correlation functions of
$\langle{\mathbf{t_i}}\cdot{\mathbf{t_j}}\rangle,
\langle{\mathbf{b_i}}\cdot{\mathbf{b_j}}\rangle,
\langle{\mathbf{n_i}}\cdot{\mathbf{n_j}}\rangle$. Thus one calculates
$\langle{\cal T}{\cal R}_k\rangle$, diagonalizes it, raises it to the
power of $j-i$, transforms it back, and performs the continuum chain
limit with $s=(j-i)b$, $l_{p}= bk/\sin(\alpha)^2$,
$a=\frac{1}{2}b\tan(\alpha)$, $(j-i)\rightarrow\infty$,
$b\rightarrow0$, i.e. $a\rightarrow0$, where $l_{p}$ is the
persistence length, $a$ is the strand separation, $b$ is the Kuhn
segment length, $0<s<L$ is the arclength, and $L$ is the contour
length. Note that within this model $\alpha$ is a fixed parameter that
determines bending characteristics of the ribbon. An exact expression
for the autocorrelation functions is obtained:
\begin{eqnarray}
  \label{eq:acft}
  \langle{\mathbf{t}}(0)\cdot{\mathbf{t}}(s)\rangle &=&
  \exp\left(-\frac{s}{l_{p}}\right)\\
  \label{eq:acfb}
  \langle{\mathbf{b}}(0)\cdot{\mathbf{b}}(s)\rangle &=&
  \exp\left(-\frac{s}{l_{p}\tan(\alpha)^2}\right)\\
  \label{eq:acfn}
  \langle{\mathbf{n}}(0)\cdot{\mathbf{n}}(s)\rangle &=&
  \exp\left(-\frac{s}{l_{p}\sin(\alpha)^2}\right).
\end{eqnarray} 
For $\alpha=\pi/2$ one recovers the usual wormlike chain result for
two dimensions. All crosscorrelation functions (the off-diagonal
elements of $\langle{\cal A}\rangle$) vanish.
Eqs.~(\ref{eq:acfb}),~(\ref{eq:acfn}) represent the persistence length
$l_{p,in}=l_{p}\tan(\alpha)^{2}$ for bending within the plane of the
ribbon and the persistence length $l_{p,out}=l_{p}\sin(\alpha)^{2}$
for bending out of the plane of the ribbon
respectively~\cite{Joanny_jp2_96}. This model was recently treated as
a twisted zig-zag fiber within the framework of a two-angle model for
studying structural properties of chromatin~\cite{Schiessel_bpj_01}.

From the tangent-tangent correlation function one can calculate the
mean squared end-to-end distance:
\begin{equation}
  \label{eq:endtoend}
  \begin{split}
    R_E^2 &= \langle({\mathbf{R}}(L)-{\mathbf{R}}(0))^2\rangle =
    \int_0^Lds_1\int_0^Lds_2 \langle{\mathbf{t}}(s_1)\cdot{\mathbf{t}}(s_2)\rangle\\
    &= 2Ll_{p}-2l_{p}^2\left(1-\exp\left(-\frac{L}{l_{p}}\right)\right).
  \end{split}
\end{equation}
Eqs.~(\ref{eq:acft}) and~(\ref{eq:endtoend}) are identical to results
for single wormlike chains~\cite{DoiEdw}. Eq.~(\ref{eq:endtoend})
interpolates between the limiting behaviors of random coils
($2Ll_{p}$) for $L\gg l_{p}$ and rigid rods ($L^2$) for $L\ll l_{p}$.


\section{Edge Stiffness}
\label{section:B}

\begin{figure}[t]  
  \begin{center}
    \includegraphics[angle=0,width=0.87\linewidth]{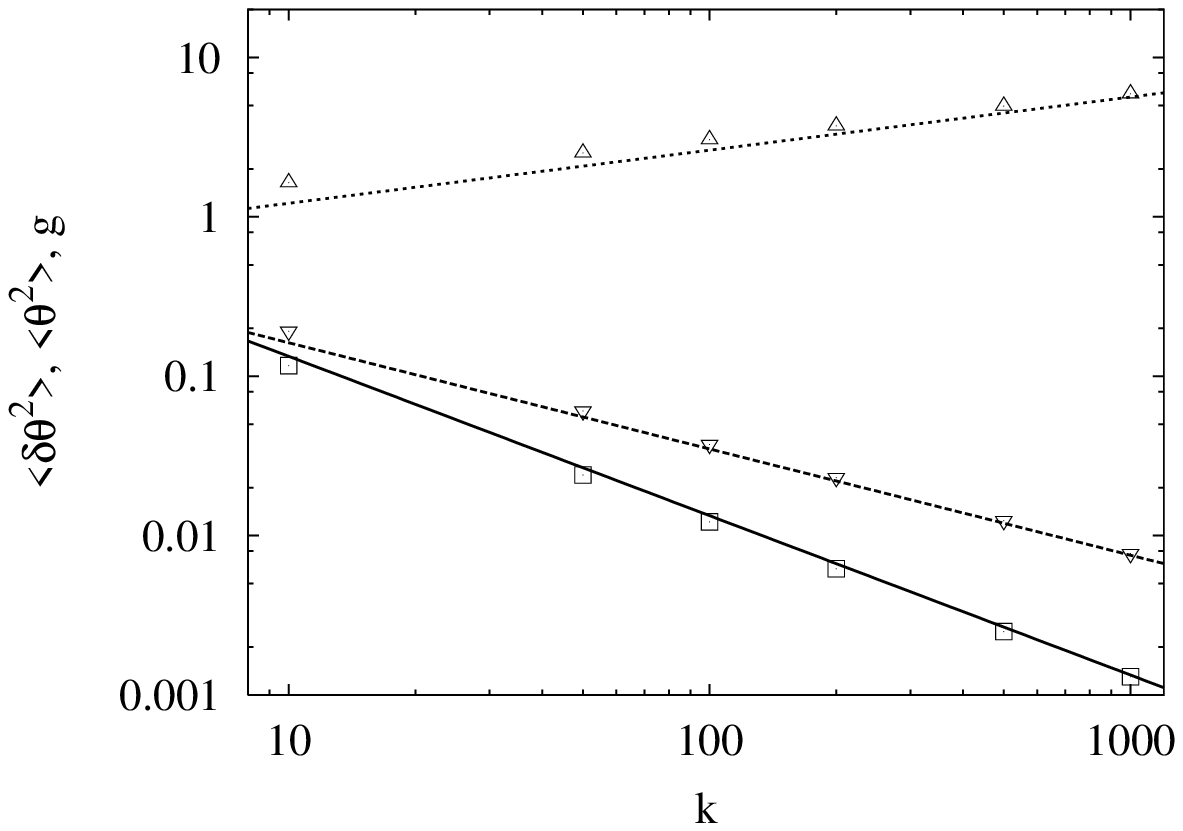}
    \includegraphics[angle=0,width=0.87\linewidth]{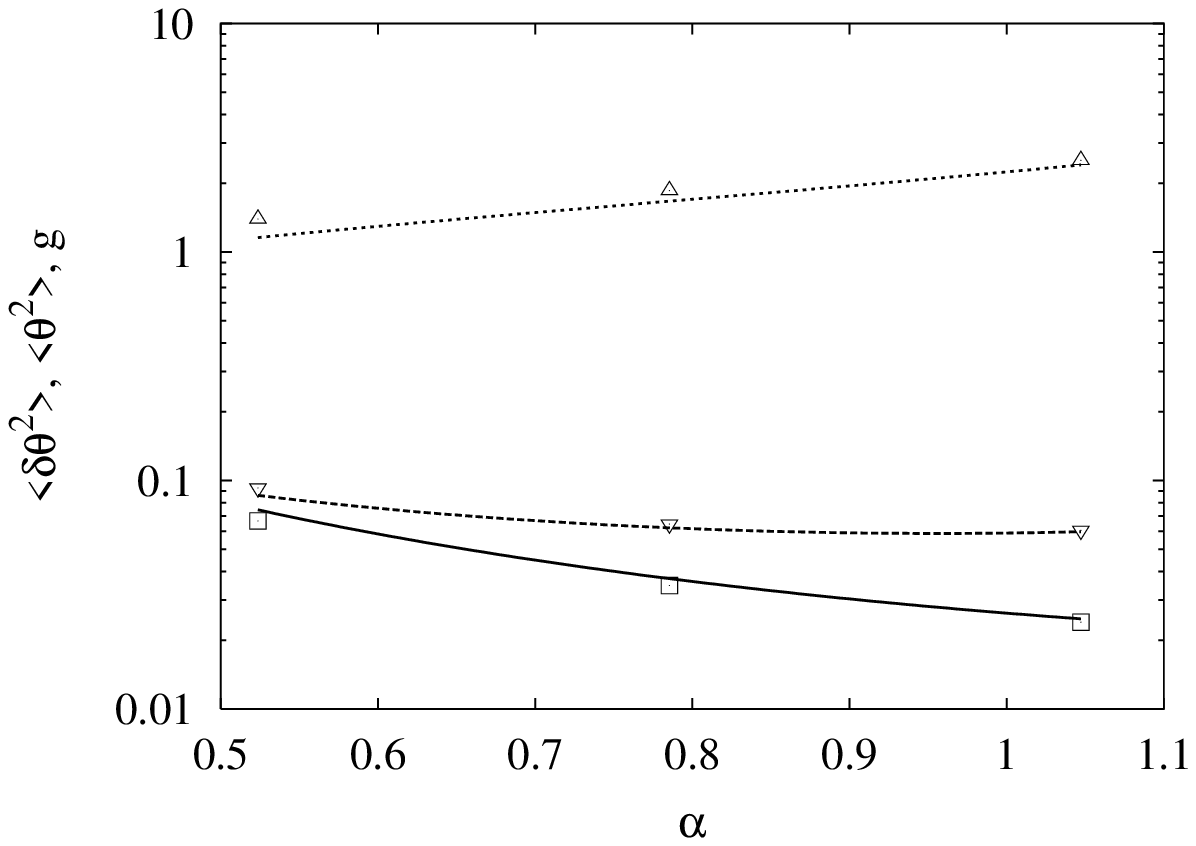}
    \includegraphics[angle=0,width=0.87\linewidth]{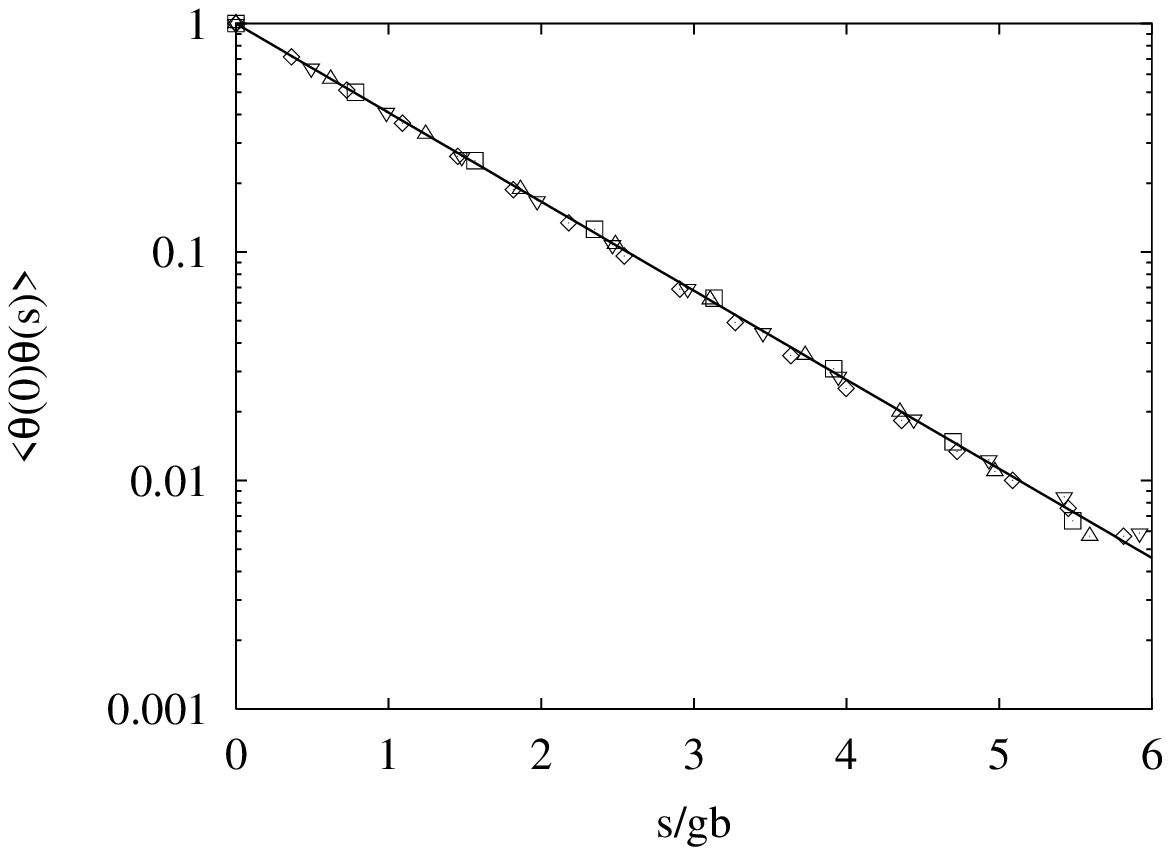}
  \end{center} 
  \caption{(a), (b) Scaling plots for $\langle\theta_i^2\rangle$
    (downward triangles), $\langle\delta\theta_i^2\rangle$ (squares),
    and g with $\alpha=\pi/3$ (upward triangles) and $k=50$
    respectively. (c) Numerical evidence for the derived expression of
    $\langle\theta_i\theta_j\rangle$. The data refer to $k=50$
    (squares), $k=100$ (circles), $k=200$ (upward triangles), $k=500$
    (downward triangles) and $\alpha=\pi/4$. We determine the correct
    prefactor $g_{fit}=0.56\pm0.05$ of $g$ from the numerical data of
    the folding angle correlation function
    $\langle\theta_i\theta_j\rangle$ which is our solely free
    parameter and use it for all following comparisons between scaling
    analysis and numerical results.}
  \label{fig:scal}
\end{figure}

In the following we present a simple scaling argument which allows us
to rationalize the behavior of the Liverpool model. Consider first the
$\delta\theta$ part of Eq.~(\ref{ht}). In the absence of other terms
the folding angles would perform a simple random walk with step length
$\langle\delta\theta_i^2\rangle=\frac{1}{k\sin(\alpha)^2}$.  The
leading term limiting the fluctuations of the folding angles around
zero is of order ${\cal O}(\theta_i^4)$. The behavior of the coupled
system can be inferred from scaling arguments similar to those used
for polymer adsorption. Consider a vanishing folding angle and follow
the chain in either direction. Up to a characteristic number of steps
$g$ the folding angles will show simple diffusion. As a consequence
the mean-squared folding angle averaged over this short segment is
$\langle\theta_i^2\rangle=g\langle\delta\theta_i^2\rangle$
corresponding to a potential energy $\frac{E_{ex}}{k_BT} \sim
g\langle\theta_i^4\rangle \sim 3g\langle\theta_i^2\rangle^2 \sim
3g^3\langle\delta\theta_i^2\rangle$. The free diffusion of the folding
angles has to stop when this potential energy is of order $k_BT$,
suggesting
\begin{eqnarray}
  \label{dt2}
  \langle\delta\theta_i^2\rangle &=& \frac{1}{k\sin(\alpha)^2}\\
  \label{gscal}
  g &\sim& \left(\frac{k\tan(\alpha)^2}{3}\right)^{\frac{1}{3}}\\
  \label{t2scal}
  \langle\theta_i^2\rangle &=& g\langle\delta\theta_i^2\rangle\\
  \label{acftheta}
  \frac{\langle\theta_i\theta_j\rangle}{\langle\theta_i^2\rangle} &=&
  \exp\left(-\frac{\left|j-i\right|}{2g}\right).  
\end{eqnarray}
Fig.~\ref{fig:scal} shows that these arguments are fully supported by
the results of our MC simulations with
$g=(0.56\pm0.05)\left(\frac{k\tan(\alpha)^2}{3}\right)^{\frac{1}{3}}$.

Using again the transfer matrix ansatz one obtains in the low
temperature limit by considering only terms of the order ${\cal
  O}(\theta_i^2)$ the following expression for
${\mathbf{t_i}}\cdot{\mathbf{t_j}}$,
${\mathbf{b_i}}\cdot{\mathbf{b_j}}$ and
${\mathbf{n_i}}\cdot{\mathbf{n_j}}$:
\begin{eqnarray}
  \label{tt}
  {\mathbf{t_i}}\cdot{\mathbf{t_j}}  &=& 1 -
  \frac{\sin(\alpha)^2}{2}\left(\sum_{k=i/2}^{j/2}\delta\theta_{2k}\right)^2\\
  \label{bb}
  {\mathbf{b_i}}\cdot{\mathbf{b_j}}  &=& 1 -
  \frac{\cos(\alpha)^2}{2}\left(\sum_{k=i}^{j-1}\theta_k^2 +
    2\sum_{k=i}^{j-1}\sum_{k'=k+1}^{j-1}\theta_k\theta_{k'}\right)\nonumber\\
  \\
  \label{nn}
  {\mathbf{n_i}}\cdot{\mathbf{n_j}}  &=& 1 - \frac{1}{2}
  \sum_{k=i/2}^{j/2} \delta\theta_{2k}^2 
  +
  \cos(\alpha)^2\sum_{k=i}^{j-1}\sum_{k'=k+1}^{j-1}\theta_k\theta_{k'}\nonumber\\.
\end{eqnarray}
\begin{figure}[t]  
  \begin{center}
    \includegraphics[angle=0,width=0.99\linewidth]{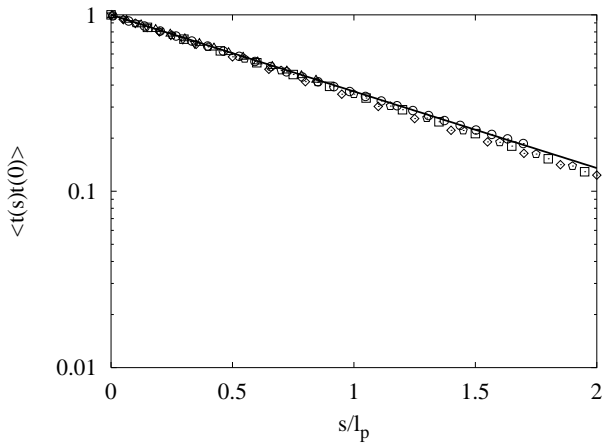}
  \end{center} 
  \caption{Comparison of MC-data and analytical results (solid line)
    for the autocorrelation function of the tangent vectors with
    $k=50$ (squares), $k=100$ (circles), $k=200$ (upward triangles),
    $k=500$ (downward triangles) and $\alpha=\pi/3$, and $k=50$ and
    $\alpha=\pi/4$ (diamonds), $\alpha=\pi/6$ (pentagons).}
  \label{fig:acf_tt}
\end{figure}
Note that $i,j$ are either odd or even depending on which strand is
under consideration. Without loss of generality we choose $i,j$ to be
even. First of all we use that
$\langle{\mathbf{t}}(0)\cdot{\mathbf{t}}(s)\rangle$ has to interpolate
between $1$ for $s=0$ and $0$ for $s\rightarrow\infty$ and that the
right hand side of Eq.~(\ref{tt}) is the Taylor expansion up to first
order of the exponential function
$\exp\left(\frac{\sin(\alpha)^2}{2}\left(\sum_{k=i/2}^{j/2}\delta\theta_{2k}\right)^2\right)$.
Substituting then $s=2|j-i|b$ and $l_{p}=4bk$, performing the
continuum chain limit with $b\rightarrow0$ and
$\alpha\rightarrow\pi/2$ respectively, i.e. keeping the strand
separation $a$ constant, yields the following expression for the
autocorrelation function of the tangent vectors:
\begin{equation}
  \label{acft}
  \langle{\mathbf{t}}(0)\cdot{\mathbf{t}}(s)\rangle =
  \exp\left(-\frac{s}{l_{p}}\right).
\end{equation}
Thus the mean squared end-to-end distance $R_{E}^2$ becomes identical
to Eq.~(\ref{eq:endtoend}). Eq.~(\ref{acft}) is confirmed by our MC
simulation data shown in Fig.~\ref{fig:acf_tt}.

\begin{figure}[t]  
  \begin{center}
    \includegraphics[angle=0,width=0.99\linewidth]{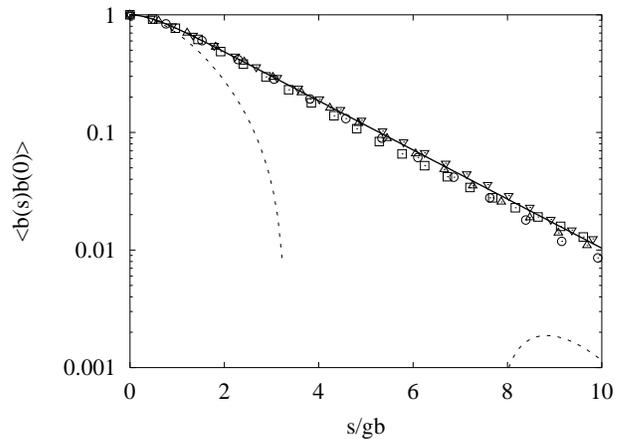}
    \includegraphics[angle=0,width=0.99\linewidth]{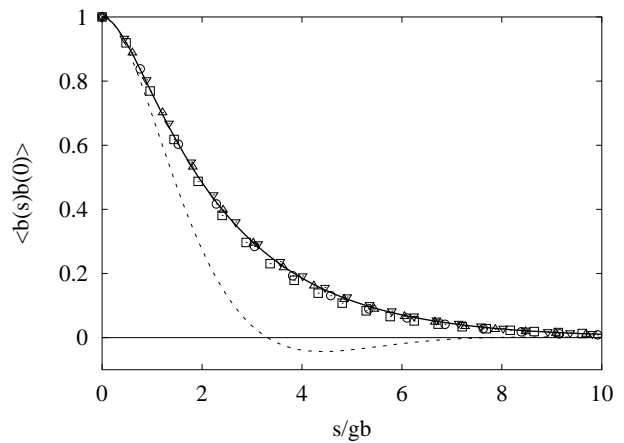}
  \end{center} 
  \caption{(a) Autocorrelation function of the bond-directors with
    $k=50$ (squares), $k=100$ (circles), $k=200$ (upward triangles),
    $k=500$ (downward triangles) and $\alpha=\pi/3$. The data show the
    predicted functional (solid line) form for
    $\langle{\mathbf{b}}(0)\cdot{\mathbf{b}}(s)\rangle$ of
    Eq.~(\ref{acfb}). In order to check the scaling argument of
    Eq.~(\ref{acfb}) we determined the correct prefactor
    $g_{fit}=0.56\pm0.05$ of $g$ with the help of the numerical data
    of $\langle\theta_i\theta_j\rangle$ (see Fig.~\ref{fig:scal}) and
    put it in Eq.~(\ref{acfb}). The agreement is excellent. The dashed
    line which oscillates is the predicted functional form of
    Liverpool {\em et
      al}~\cite{Liverpool_prl_98,LiverpoolGolestian_pre_00}.\newline
    (b) Comparison of our simulation data with the analytical result
    of Liverpool {\em et al} (dashed line). The predicted oscillation
    and resultant pitch is not recovered. But we find the same scaling
    behavior of the helical persistence length with $l_b=gb\sim
    l_p^{\frac{1}{3}}a^{\frac{2}{3}}$. It is also striking that the
    predicted functional form of Liverpool {\em et al} is in very good
    agreement with our numerical data within one helical persistence
    length $l_b$.}
  \label{fig:acf_bb}
\end{figure}

To get an idea of the structural properties characterized by the
autocorrelation function of the bond-directors
$\langle{\mathbf{b_i}}\cdot{\mathbf{b_j}}\rangle$ we calculate the
mean squared twist $\langle Tw(i,j)^2\rangle$ of the ribbon.
Following the definition of the local twist rate $\tau_i$ of
Eq.~(\ref{tau}) the total twist between two triangles of index $i$ and
$j$ is just the sum of the local twist angles determined by the
projections of the normal vector of the $i$th triangle onto the
bond-director of the ($i+1$)th triangle, that is
\begin{equation}
\label{tw}
  Tw(i,j) = \frac{1}{2\pi} \sum_{k=i}^{j-1}{\mathbf{n_{i}}}\cdot{\mathbf{b_{i+1}}} =
  \frac{\cos(\alpha)}{2\pi}\sum_{k=i}^{j-1}\theta_k.
\end{equation}
Comparing Eq.~(\ref{bb}) and~(\ref{tw}) we find for small twist angles
\begin{equation}
  \label{eq:twbb}
  \langle{\mathbf{b_i}}\cdot{\mathbf{b_j}}\rangle = 1 - 2\pi^2 \langle Tw(i,j)^2\rangle.
\end{equation}
Hence the autocorrelation function of the bond-directors can be seen as 
a measure for the local twist structure of the ribbon.

\begin{figure}[t]  
  \begin{center}
    \includegraphics[angle=0,width=0.99\linewidth]{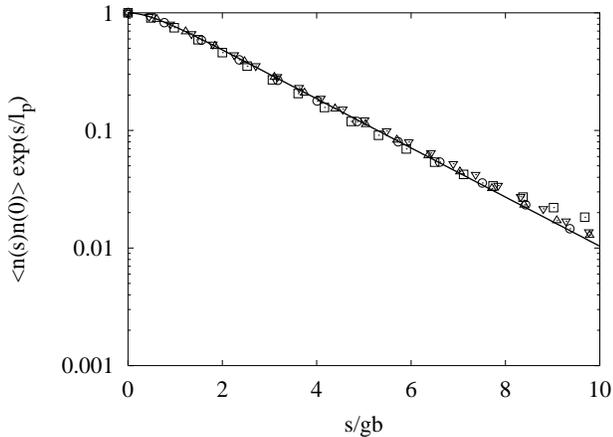}
  \end{center} 
  \caption{Autocorrelation function of the normal vectors with
    $k=50$ (squares), $k=100$ (circles), $k=200$ (upward triangles),
    $k=500$ (downward triangles) and $\alpha=\pi/3$. We divided out of
    the normal vector correlation function (solid line) the tangent
    correlation function
    $\langle{\mathbf{t}}(0)\cdot{\mathbf{t}}(s)\rangle$ (see
    Eq.~(\ref{acfn})) so that one should regain the same exponential
    decay as for $\langle{\mathbf{b}}(0)\cdot{\mathbf{b}}(s)\rangle$
    which is in agreement with the numerical data.}
  \label{fig:acf_nn}
\end{figure}
In contrast to the plaquette stiffness model the angles $\theta_i$ are
correlated (see Eq.~(\ref{acftheta})). Therefore the double summation
over $\langle\theta_i\theta_j\rangle$ in Eq.~(\ref{bb}) yields an
analogous result as it is obtained in the calculation of the mean
squared end-to-end distance of the wormlike chain model. Using the
derived scaling expressions of Eqs.~(\ref{gscal}) and~(\ref{t2scal}), the
same substitutions as in Eq.~(\ref{acft}), and performing the
continuum chain limit one obtains the following relationship
for the autocorrelation function of the bond-directors:
\begin{equation}
  \label{acfb}
  \langle{\mathbf{b}}(0)\cdot{\mathbf{b}}(s)\rangle =
  \exp\left(-2\pi^2\langle Tw(0,s)^2\rangle\right)
\end{equation}
with
\begin{equation}
  \label{TwB}
  \langle Tw(0,s)^2\rangle =
  \frac{1}{3\pi^2}\left(\frac{s}{gb}
    -
    2\left(1-\exp\left(-\frac{s}{2gb}\right)\right)\right)
\end{equation}
and $gb = g_{fit}^{1/3} l_{p}^{\frac{1}{3}}a^{\frac{2}{3}}/3^{1/3}$,
where $g_{fit}=0.56\pm0.05$ is the fitted prefactor for the scaling
function $g$. $a$ represents the strand separation of the ribbon which
is given by $a=|{\mathbf{b_i}}|=\frac{1}{2}b\tan(\alpha)$. Hence we
observe two length scales influencing the local twist structure of the
ribbon: on the one hand the single strand persistence length $l_{p}$
and on the other hand the strand separation $a$. The predicted scaling
behavior of $\langle{\mathbf{b}}(0)\cdot{\mathbf{b}}(s)\rangle$ can be
observed in the simulation data as it is shown in
Fig~(\ref{fig:acf_bb}). Note that
$\langle{\mathbf{b}}(0)\cdot{\mathbf{b}}(s)\rangle$ as well as all
other calculated observables within this model is independent of the
geometry of the triangles in contrast to the previous model where
$\alpha$ was a fixed parameter.

Eq.~(\ref{nn}) can be evaluated in the same manner resulting in
\begin{equation}
  \label{acfn}
  \begin{split}
    \langle{\mathbf{n}}(0)\cdot{\mathbf{n}}(s)\rangle &=
    \langle{\mathbf{t}}(0)\cdot{\mathbf{t}}(s)\rangle
    \langle{\mathbf{b}}(0)\cdot{\mathbf{b}}(s)\rangle\\
    &=
    \exp\left(-\frac{s}{l_{p}}-2\pi^2\langle Tw(0,s)^2\rangle\right).
  \end{split}
\end{equation}
Eq.(\ref{acfb}) shows that the autocorrelation function of the normal
vectors is the product of
$\langle{\mathbf{t}}(0)\cdot{\mathbf{t}}(s)\rangle$ and
$\langle{\mathbf{b}}(0)\cdot{\mathbf{b}}(s)\rangle$. For very stiff
chains, the tangent correlation function gives just small corrections
to the normal vector correlation function. Therefore one can
interpret Eq.(\ref{acfb}) as the rigid rod limit of Eq.(\ref{acfn}).

\begin{figure}[t]  
  \begin{center}
    \includegraphics[angle=0,width=0.99\linewidth]{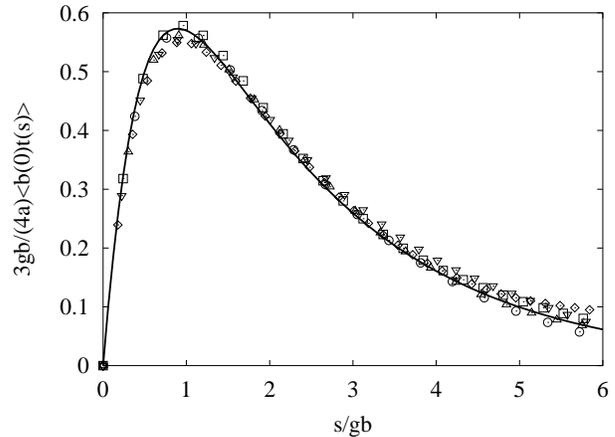}
  \end{center} 
  \caption{Crosscorrelation function of the bond-directors and the tangent vectors with
    $k=50$ (squares), $k=100$ (circles), $k=200$ (upward triangles),
    $k=500$ (downward triangles), $k=1000$ (diamonds) and
    $\alpha=\pi/3$. The data validate the predicted functional form
    (solid line) for
    $\langle{\mathbf{b}}(0)\cdot{\mathbf{t}}(s)\rangle$ of
    Eq.~(\ref{ccf}).}
  \label{fig:ccf}
\end{figure}

Other important structural properties of the ribbon can be extracted
out of the crosscorrelation functions.
$\langle{\mathbf{n}}(0)\cdot{\mathbf{t}}(s)\rangle$ and
$\langle{\mathbf{n}}(0)\cdot{\mathbf{b}}(s)\rangle$ describe the mean
curvature and mean twist respectively and vanish in both models for
symmetry reasons. For
$\langle{\mathbf{b}}(0)\cdot{\mathbf{t}}(s)\rangle$ we empirically
observe the following relationship:
\begin{equation}
  \label{ccf}
  \langle{\mathbf{b}}(0)\cdot{\mathbf{t}}(s)\rangle =
  (2\pi)^2a\frac{d}{ds}\langle Tw(0,s)^2\rangle e^{-(2\pi)^2\langle Tw(0,s)^2\rangle}.
\end{equation}
Eq.~(\ref{ccf}) can be understood qualitatively in the following way.
Due to the anisotropic rigidity of the ribbon the scalar product
${\mathbf{b}}(0)\cdot{\mathbf{t}}(s)$ is only non-zero if the chain is
bent and twisted simultaneously. In case the ribbon is either solely
bent or solely twisted the bond-directors are always perpendicular to
the tangent vectors and the scalar product
${\mathbf{b}}(0)\cdot{\mathbf{t}}(s)$ vanishes for all $s$. The rate
of mean twist of one helical persistence length $l_b=gb$ which
defines the size of the locally existing helical structures can be
calculated with Eq.~\ref{TwB} yielding $\sqrt{\langle
  Tw(0,bg)^2\rangle}\approx\pm\frac{1}{16}$. This corresponds to a
typical twist angle of $\Psi=\frac{\pi}{8}$ using $Tw=2\pi\Psi$.
Within $l_b$ the twist rate is determined by the derivative of the
mean squared twist $\frac{d}{ds}\langle Tw(0,s)^2\rangle$ which gives
rise to the increasing correlation function
$\langle{\mathbf{b}}(0)\cdot{\mathbf{t}}(s)\rangle$ up to the maximum
value at $l_b=gb$. For larger internal distances of the chain the rate
of mean twist is a random sequence of $\pm\frac{1}{16}$ so that the
crosscorrelation function has to vanish and therefore decreases
exponentially with $\exp\left(-(2\pi)^2\langle
  Tw(0,s)^2\rangle\right)$. Fig.~(\ref{fig:ccf}) compares
Eq.~(\ref{ccf}) with our numerical data. The agreement is excellent.


\section{Behavior under compression: Euler Buckling vs. Kinks}
\label{sec:buck}

As discussed in section \ref{section:model} the edge stiffness model
includes local twist correlations at least on small length scales as a
consequence of the correlation of the folding angles
$\{\theta_i\}$. In order to understand and to quantify the effects
arising from the local twist we measured the probability distribution
functions of the folding angles, of the twist, and of the
end-to-end-distance for different rigidities and compared the latter
with the usual wormlike chain model to see which differences occur.

If there is a preference for kinking one can enforce this property by
applying an additional constant force ${\mathbf F}_{buck}=f{\mathbf
  R}_E/R_E$ which compresses the ribbon. In addition the change in
the end-to-end distance $R_E$ caused by the buckling force should
affect the twist distribution function $P(Tw)$ if $R_E$ and $Tw$ are
coupled.

\begin{figure}[t]  
  \begin{center}
    \includegraphics[angle=0,width=0.99\linewidth]{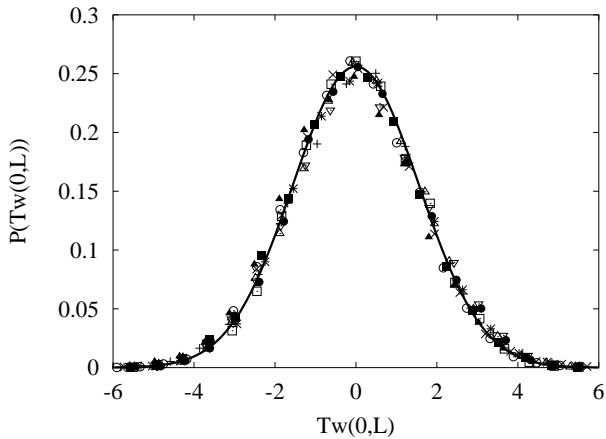}
  \end{center} 
  \caption{Comparison of the probability distribution functions of the
    total twist of the ribbon for both models with
    $f=f=\{0,0.01,0.02,...,0.09\}$ and $l_p=L=400$ with the scaling
    analysis for $f=0$. One recovers the same Gaussian shape for all
    values of $f$.}
  \label{fig:twist}
\end{figure}
For small forces we calculate the change of twist under the influence
of the external force $F_{buck}=f$ within the framework of linear
response theory:
\begin{equation}
  \begin{split}
    &\langle \Delta Tw(0,L)^2\rangle = \langle Tw(0,L)^2\rangle - \langle
    Tw(0,L)^2\rangle_{f=0}\\
    &= -\beta f\left(\langle
      R_ETw(0,L)^2\rangle_{f=0} - \langle R_E\rangle_{f=0}\langle
      Tw(0,L)^2\rangle_{f=0}\right)
  \end{split}
\end{equation}
with $\beta=1/k_BT$. This predicts a change of the mean squared twist
of the chain if a twist-stretch coupling determined by $\langle
R_ETw(0,L)^2\rangle_{f=0}$ exists. Note that $\langle
R_ETw(0,L)\rangle_{f=0}$ vanishes due to symmetry reasons. The
evaluation of our numerical data yields that $\langle
R_ETw(0,L)^2\rangle$ is uncorrelated, too. To quantify if higher order
terms in $f$ contribute to a change of $\langle Tw(0,L)^2\rangle$ we
carried out several simulation runs with varying force strengths
$f=\{0,0.01,0.02,...,0.09\}$ corresponding to
$\frac{R_E(f)}{R_E(0)}=\{1,0.95,0.87,0.71,0.51,0.36,0.26,0.21,0.17,0.15\}$.

Fig.~\ref{fig:twist} shows the same Gaussian shape for all measured
probability distribution functions of the total twist of the ribbon
$P(Tw(0,L),f)$. This implicates that there is no twist-stretch
coupling inherent in the system. The same is valid for the
distribution function of the folding angles.

\begin{figure}[t]  
  \begin{center}
    \includegraphics[angle=0,width=0.99\linewidth]{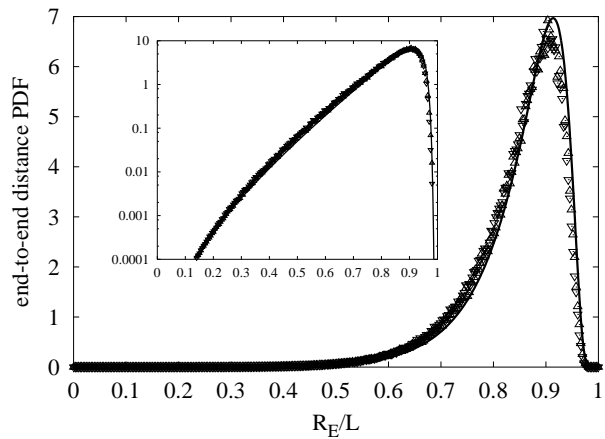}
  \end{center} 
  \caption{Probability distribution functions of the end-to-end
    distance of the edge stiffness model for different discretizations
    ($N=800$ upward triangles, $N=600$ downward triangles) with
    $\alpha=\frac{\pi}{4}$ and $l_{p}=L=400$ calculated with the help
    of the multiple histogram method~\cite{FeSw_prl_88} and the usual
    wormlike chain model (solid line). The PDF of the wormlike chain
    model is calculated with the derived analytical expression of
    Wilhelm and Frey~{\cite{wilhelm_prl_96}}.}
  \label{fig:rehist}
\end{figure}
\begin{figure}[t]  
  \begin{center}
    \includegraphics[angle=0,width=0.99\linewidth]{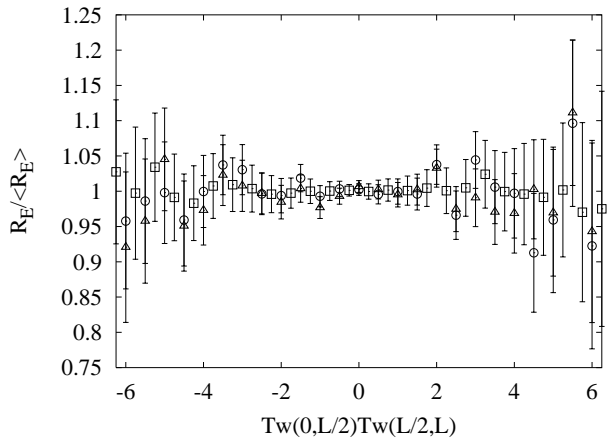}
  \end{center} 
  \caption{End-to-end distance $R_E$ as a function of the product of
    the twist left and right of the center of the chain
    $Tw(0,\frac{L}{2})Tw(\frac{L}{2},L)$, which is a measure for
    unlike (negative sign) and like (positive sign) twists meeting at
    the center, and as a function of the applied buckling force with
    $f=0\,\text{(squares)},0.03\,\text{(circles)},0.06\,\text{(triangles)}$
    and $l_p=0.5L=200$. $R_E$ refers to the average of one interval of
    $Tw(0,\frac{L}{2})Tw(\frac{L}{2},L)$ and $\langle R_E\rangle$
    refers to the mean value of all sampled end-to-end distances. One
    does not find an asymmetry between end-to-end distances for like
    and unlike twists meeting at the center. The larger fluctuations
    for larger values of $Tw(0,\frac{L}{2})Tw(\frac{L}{2},L)$ are the
    result of a poorer sampling rate.}
  \label{fig:kink}
\end{figure}

Moreover we measured the probability distribution function $P(R_E,f)$
of the end-to-end distance $R_E$ for all applied forces $f$. Using the
multiple histogram method developed by Ferrenberg and
Swendsen~\cite{FeSw_prl_88} one can then recombine all measured
histograms with a reweighting procedure to a single probability
distribution function $P(R_E)$ with overall very good statistics.
Fig.~\ref{fig:rehist} shows $P(R_E)$ for ${\cal H}_{tt}$ and the
wormlike chain model. Quite contrary to a shift to noticeably shorter
end-to-end distances $R_E$ as one would expect for the above described
phenomena of kinks one just recovers the usual wormlike chain
behavior. This indicates that the ribbon just bends under the external
force in contradiction to a kink-rod structure. Another quantity which
is sensitive to the presence of kinks is a three-point correlation
function of the end-to-end distance $R_E$ and the twist to the left
$Tw(0,\frac{L}{2})$, and to the right $Tw(\frac{L}{2},L)$ of the
center of the chain. Due to the buckling force the center of the chain
is labeled which means that a kink is detected if the end-to-end
distances with $Tw(0,\frac{L}{2})Tw(\frac{L}{2},L)<0$ (unlike twists
meeting at the center) are smaller than the end-to-end distances with
$Tw(0,\frac{L}{2})Tw(\frac{L}{2},L)>0$ (like twists meeting at the
center). Fig.~\ref{fig:kink} shows the mean end-to-end distance
depending on the value of $Tw(0,\frac{L}{2})Tw(\frac{L}{2},L)$ for
$l_p=200$, $L=400$, and $f=0$, $f=0.03$, $f=0.06$. We do not find an
asymmetry between like and unlike twists meeting at the center as it
would support the prediction of kinks made by Liverpool {\em et
  al}~\cite{Liverpool_prl_98,LiverpoolGolestian_pre_00}.

\section{Summary}

We have reinvestigated the mechanical properties of the model
introduced by Liverpool {\em et
  al}~\cite{Liverpool_prl_98,LiverpoolGolestian_pre_00} of a
double-stranded semiflexible polymer and rationalized the results of
our MC simulations with the help of a simple scaling argument. We
recover the predicted simple exponential decay of the tangent-tangent
correlation function with the single strand persistence length $l_p$
and that ${\mathbf t}(0)\cdot{\mathbf t}(s)$ is independent of the
separation $a$ of the two strands, which is in addition to $l_p$ the
other relevant length scale in the problem. Also in agreement with
Ref.~\cite{Liverpool_prl_98,LiverpoolGolestian_pre_00} we find that
the helical persistence length $l_b$ and the helical pitch $P$ scale
with $l_p^{1/3}a^{2/3}$. Qualitatively, one would expect to see
oscillations in the bond-director correlation function, if $P\le l_b$.
This can be understood by calculating the rate of mean twist within
$l_b=gb$, i.e.  $\sqrt{\langle Tw(0,gb)^2\rangle}$. If the mean twist
rate exceeds $\pi$ an oscillatory behavior has to be observed. But our
calculation gives a twist rate within $l_b=gb$ of approximately
$\pm1/16$. For larger distances of the chain the rate of mean twist is
just given by a random sequence of $\pm1/16$ and thus cannot account
for an oscillatory behavior of
$\langle{\mathbf{b}}(0)\cdot{\mathbf{b}}(s)\rangle$. Liverpool {\em et
  al} predict $P=l_b$, while our analysis indicates $P=16l_b$ as it is
demonstrated in Fig.~\ref{fig:acf_bb}~(b). The authors claimed support
from their own simulations, but failed to provide a quantitative
comparison between their numerical and analytical results. In fact the
presented oscillations seem to be ordinary fluctuations within the
statistical errors. But as can be seen in Fig.~\ref{fig:acf_bb}~(b)
the predicted functional form for the bond-director autocorrelation
function is in very good agreement with our numerical data as well as
with our scaling results within one helical persistence length
$l_b=gb$.

Moreover our simulation results with applied constant buckling forces
do not provide any evidence of the predicted tendency of kinking or
the claimed twist-stretch coupling. Thus contrary to the claim made in
Ref.~\cite{Liverpool_prl_98,LiverpoolGolestian_pre_00} the local
twist structure does not suffice to explain experimental observations
such as the twist-stretch coupling~\cite{Strick_sci_96,Strick_bpj_98}
and the kink-rod structures~\cite{Kaes_epl_93} of helical
double-stranded molecules. These features require the inclusion of a
spontaneous twist incorporated by an additional term in the
Hamiltonian, e.g. ${\cal
  H}_{Tw}=k_{Tw}\sum_i\left(\sum_{j=i}^{i+1}{\mathbf{n}_j}\cdot{\mathbf{b}_{j+1}}-\theta_{sp,i}\right)^2$,
~\cite{Rabin_condmat_01,Rabin_pre_00,Rabin_prl_00,Marko_epl_97,Kamien_epl_97,Haijun_bpj_00,Zhou_pre_00,Haijun_prl_99}.\newline

\section{Acknowledgments}

The authors would like to thank B. D\"unweg and H.J. Limbach, T.B.
Liverpool, K. Kremer and H. Schiessel for numerous helpful
discussions. Financial support from the DFG within the Emmy-Noether
Programm is greatfully acknowledged.

\bibliographystyle{/people/thnfs/homes/mergell/latex/Ribbon/phaip}

\end{document}